\documentclass[12pt,preprint]{aastex}

\shortauthors{White, Becker, Fan, Strauss}
\shorttitle{Probing Ionization at $z>6$}

\slugcomment{To appear in the Astronomical Journal, July 2003}

\newcommand{\qa}{SDSS J1030$+$0524}
\newcommand{\qb}{SDSS J1148$+$5251}
\newcommand{\lya}{Ly$\,\alpha$}
\newcommand{\lyb}{Ly$\,\beta$}
\newcommand{\lyg}{Ly$\,\gamma$}
\newcommand{\nhrat}{n_{\rm HI}/n_{\rm H}}
\newcommand{\nhratover}{n_{\rm HI} \over n_{\rm H}}
\newcommand{\taua}{\tau(\hbox{\lya})}
\newcommand{\taub}{\tau(\hbox{\lyb})}

\begin{document}
\title{Probing the Ionization State of the Universe at $z>6$ \altaffilmark{1}}
\author{
Richard~L.~White\altaffilmark{2,3},
Robert~H.~Becker\altaffilmark{4,5},
Xiaohui Fan\altaffilmark{6},
\&
Michael A. Strauss\altaffilmark{7}}
\email{rlw@stsci.edu}

\altaffiltext{1}{Based on observations obtained with the W.~M.~Keck
Observatory, which is jointly operated by the California Institute of
Technology and the University of California.}
\altaffiltext{2}{Space Telescope Science Institute, Baltimore, MD 21218}
\altaffiltext{3}{Dept. of Physics \& Astronomy, Johns Hopkins University}
\altaffiltext{4}{Physics Dept., University of California, Davis, CA 95616}
\altaffiltext{5}{IGPP/Lawrence Livermore National Laboratory}
\altaffiltext{6}{Steward Observatory, University of Arizona, Tucson, AZ 85721}
\altaffiltext{7}{Princeton University Observatory,
Princeton, NJ 08544}

\begin{abstract}

We present high signal-to-noise ratio Keck ESI spectra of the two
quasars known to have Gunn-Peterson absorption troughs, \qa\ ($z=6.28$)
and \qb\ ($z=6.37$).   The \lya\ and \lyb\ troughs for \qa\ are very
black and show no evidence for any emission over a redshift interval of
$\sim0.2$ starting at $z=6$.  On the other hand, \qb\ shows a number of
emission peaks in the \lyb\ Gunn-Peterson trough along with a single
weak peak in the \lya\ trough.  The \lya\ emission has corresponding
\lyb\ emission, suggesting that it is indeed a region of lower optical
depth in the intergalactic medium at $z=6.08$.

The stronger \lyb\ peaks in the spectrum of \qb\ could conceivably
also be the result of ``leaks'' in the IGM, but we suggest that
they are instead \lya\ emission from an intervening galaxy at
$z=4.9$.  This hypothesis gains credence from a strong complex of
\ion{C}{4} absorption at the same redshift and from the detection
of continuum emission in the \lya\ trough at the expected brightness.
If this proposal is correct, the quasar light has probably been
magnified through gravitational lensing by the intervening galaxy.
The Str\"omgren sphere observed in the absorption spectrum of \qb\ is
significantly smaller than expected based on its brightness, which
is consistent with the hypothesis that the quasar is lensed.

If our argument for lensing is correct, the optical depths derived
from the troughs of \qb\ are only lower limits (albeit still
quite strong, with $\taua > 16$ inferred from the \lyb\ trough.)
The \lyb\ absorption trough of \qa\ gives the single best measurement
of the IGM transmission at $z>6$, with an inferred optical depth
$\taua>22$.

\end{abstract}

\section{Introduction}

The discovery of quasars at redshifts greater than 6 from the Sloan
Digital Sky Survey (SDSS; Fan et al.\ 2001, Fan et al.\ 2003) has
opened a window for spectroscopic exploration of the high redshift
intergalactic medium.  The detection of broad, black, \lya\ absorption
troughs, as predicted by Gunn \& Peterson (1965; hereafter GP), indicates that
we are beginning to probe the era when hydrogen was reionized by
an early generation of stars.  Becker et al.\ (2001) detected the
first complete GP trough in the spectrum of
SDSSp~J103027.10$+$052455.0 ($z=6.28$; hereafter \qa).  They compared
the absorption to that seen in the spectra of lower redshift quasars
and concluded that the fraction of neutral hydrogen in the IGM increases
sharply at $z\sim6$.  Fan et al.\ (2003) report that the most distant
known quasar, SDSS~J114816.64$+$525150.3 ($z=6.37$\footnote{This is
a new redshift, different from that given in Fan et al.\ (2003), based
on the spectrum presented in this paper.  See \S\ref{section-redshift}
for a discussion of the $z=6.41$ redshift derived by
Willott, McLure \& Jarvis (2003) for this object.};
hereafter \qb),
also shows a complete GP trough in its spectrum.
Djorgovski, Castro, Stern \& Mahabal (2001) argued that the spectra
of even lower redshift quasars show indications of neutral regions
in the IGM.

There ensued a spate of theoretical papers discussing the details of
how reionization takes place.  But the most exciting new observational
result is undoubtedly the discovery by the Wilkinson Microwave
Anisotropy Probe (WMAP) that the microwave background has traversed a
moderate optical depth ($\tau=0.17\pm0.04$) electron scattering medium,
implying that the universe must have been ionized at $z\sim10$--20
(Kogut et al.\ 2003).  At first glance this appears inconsistent with
reionization at $z\sim6$, but since a relatively small neutral fraction
($\sim1$\%) would suffice to produce black GP troughs, the results are
not contradictory.  There is mounting evidence that the ionization
history of the intergalactic medium (IGM) was complex.  Wyithe \& Loeb
(2003) suggested that the hydrogen in the IGM could have been reionized
twice, and Cen (2003) argued that two phases of reionization are likely
under a wide range of conditions.  In this scenario, the first
reionization occurs at $z\sim15$, driven by the formation of massive,
zero-metallicity Population~III stars (Venkatesan, Tumlinson \& Shull
2003; Mackey, Bromm \& Hernquist 2003).  The ionizing radiation from
massive stars is dramatically reduced, however, when Pop~III supernovae
seed the gas with metals (due to a reduction in the formation of
massive stars, increased line-blanketing in hot star atmospheres, and
increased mass loss rates from both hot stars and red giants.)
Consequently the IGM recombines and remains relatively neutral, $\nhrat
> 0.1$, until an increasing Pop~II massive star population and a
declining IGM density allow the second reionization to occur at
$z\sim6$. This neutral fraction is sufficient to produce the observed
high optical depths in the \lya\ GP troughs of high redshift quasars.

To advance our understanding of the ionization history of
the universe, we need high resolution, high signal-to-noise
ratio spectra of the IGM absorption at $z>6$.
Recently several groups have discovered \lya-emitting galaxies at
$z\sim6.5$ (Hu et al.\ 2002, Kodaira et al.\ 2003).  These objects
are interesting and relevant as direct probes of star and galaxy
formation at $z>6$, and the mere detectability of their \lya\ emission
is an important clue to the ionization state of the IGM (Haiman 2002).  However,
their continua are far too faint (even redward of \lya) to make
them useful for studies of the intervening IGM.  Only high-redshift
quasars offer the possibility of acquiring high quality information
on the state of the IGM at high redshift.

In this paper we present high quality spectra of the only two quasars
known to have GP absorption troughs\footnote{SDSS J1048$+$4637 at
$z=6.23$ (Fan et al.\ 2003) could also have a GP trough, but our recent
spectra show that it has strong, intrinsic, broad absorption lines
that make it less useful for studies of the IGM at $z>6$ (Fan et
al.\ 2003b).},
\qa\ ($z=6.28$) and \qb\ ($z=6.37$), taken with the ESI spectrograph on
the Keck Telescope.  The following sections describe the observations
and data reduction (\S2), discuss the analysis methods used (\S3),
present the properties of the IGM absorption in the spectra (\S4), and
discuss the results (\S5).  Appendix~\ref{section-appendix}
explores the effect of light travel time on the observed
expansion of ionization fronts around high-redshift quasars.

\section{Observations}

Spectra of \qa\ and \qb\ were taken with the Keck Echellette
Spectrograph and Imager (ESI; Sheinis et al.\ 2002) between January
2002 and February 2003.  ESI covers the wavelength range
4000--11000~\AA\ with 10 Echelle orders having a constant resolution of
$R = \lambda/\Delta\lambda \sim 10000$ ($\sim30$~km/s).  The data were
taken with the slit at the parallactic angle in mostly good but
variable seeing (from 0.6\arcsec\ to $>1$\arcsec, with mean seeing
around 0.8\arcsec).  The sky
transparency ranged from photometric to light cirrus.  A log of the
observations is given in Table~1.

\begin{deluxetable}{ccccc}
\tablecolumns{5}
\tablewidth{0pc}
\tablecaption{Keck ESI Observation Log}
\tablehead{
\colhead{Source} & \colhead{Date} & \colhead{Slit Width} & \colhead{Number of} & \colhead{Total Exposure} \\
\colhead{} & \colhead{yyyy/mm/dd} & \colhead{(\arcsec)} & \colhead{Exposures} & \colhead{(hr)} \\
}
\startdata
\qa & 2002/01/10 & 1.00 &  6 & 2.00 \\
\qa & 2002/01/11 & 1.00 &  5 & 1.67 \\
\qa & 2002/01/12 & 1.00 & 12 & 4.00 \\
\qa & 2002/01/13 & 0.75 &  2 & 1.50 \\
\qa & 2002/02/07 & 0.75 &  2 & 1.50 \\
\qb & 2002/12/04 & 1.00 &  9 & 3.00 \\
\qb & 2002/12/05 & 1.00 & 10 & 3.33 \\
\qb & 2003/02/04 & 1.00 & 14 & 4.67 \\
\enddata
\end{deluxetable}

The most challenging data reduction problem for these observations
is subtracting the sky background, including the bright OH lines that dominate
the near infrared atmospheric emission.  The sky is considerably
brighter than the object, and small errors in the sky subtraction
can easily produce offsets in the emission that either create the
appearance of light in the absorption trough or remove light that
should be present there.  To assist in the sky subtraction, we
shifted the object along the slit for the various integrations,
which helps to average over any small-scale flat-fielding, bias or dark
current variations not corrected by the standard CCD calibrations.

The sky brightness
varies as a function of time in a complex way in the near
infrared, with the atmospheric OH emission varying rapidly and
independently of other sky emission.  In addition the ESI spectrograph,
though a highly stable instrument, does undergo small changes 
over time, with the wavelengths shifting slightly from
one observation to the next.  Presumably these changes (which are
typically at the level of a few tenths of a pixel or less) are due
to minor thermal and gravity-induced flexures.

We extract a sky-subtracted spectrum from
each independent observation.  The separate spectra are then combined
using weights to optimize the signal-to-noise ratio of the result.  It
is necessary to be very careful with the weights for combining to
avoid any bias in the results.  A simple unweighted average is
safely free of bias but leads to a much noisier spectrum than an
optimally weighted sum.  On the other hand, using weights $1/\sigma^2$
with $\sigma$ derived from the noise estimates of the individual
spectra leads to a very noticeable bias toward lower flux values,
because the estimated noise is larger for brighter pixels.  In the
black absorption troughs this leads to fluxes that are systematically
negative.

The dangers of using a weighted combination for Poisson data are
well-documented (for a nice analysis, see Wheaton et al.\ 1995).
The cure for the bias is to choose weights that are completely
independent of the data.  We have combined the spectra using weights
that are derived from the sky noise in adjacent pixels (i.e.,
along the slit).  Note that
using the sky noise from the same pixel would still lead to a
small bias, because pixels that have positive sky fluctuations
(making the residual counts a little low) would be given slightly
lower weights, thus leading to a positive bias in the sum.  Using
the sky noise in neighboring pixels produces a completely unbiased
result that has a nearly optimal signal-to-noise ratio.  An overall
scale factor is determined for each spectrum to match the fluxes,
thereby correcting for variable atmospheric transparency.

Our extraction algorithm is highly customized
for ESI data, which have a number of unusual characteristics
(e.g., highly curved orders, a strongly non-linear wavelength
dispersion, and some odd notches in the sensitivity in the bluer
orders.) The images are first geometrically rectified to produce
orders with orthogonal wavelength and spatial coordinates lying
along rows and columns.  This removes the very large curvature of
the ESI Echelle orders and also corrects small variations in the
wavelength dispersion along the slit that manifest themselves as
tilted or curved sky lines.  The order curvature removal also
automatically corrects the spectral trace for differential atmospheric
refraction using the airmass of the observation; this approach is more
effective than tracing the orders directly because high-redshift
quasar spectra have long stretches with no detectable light, foiling
standard tracing algorithms.

The resampling algorithm used in the rectification is a flux-conserving
interpolation scheme that is accurate to second order in the pixel shifts (as
opposed to the more commonly used schemes that are only first
order, making them considerably more dispersive.)
The interpolation method is based on the high-order,
monotonic, flux-conserving advection schemes that are widely used
in modern hydrodynamic simulations (van Leer 1977).

\begin{figure*}
\epsscale{1.0}
\plotone{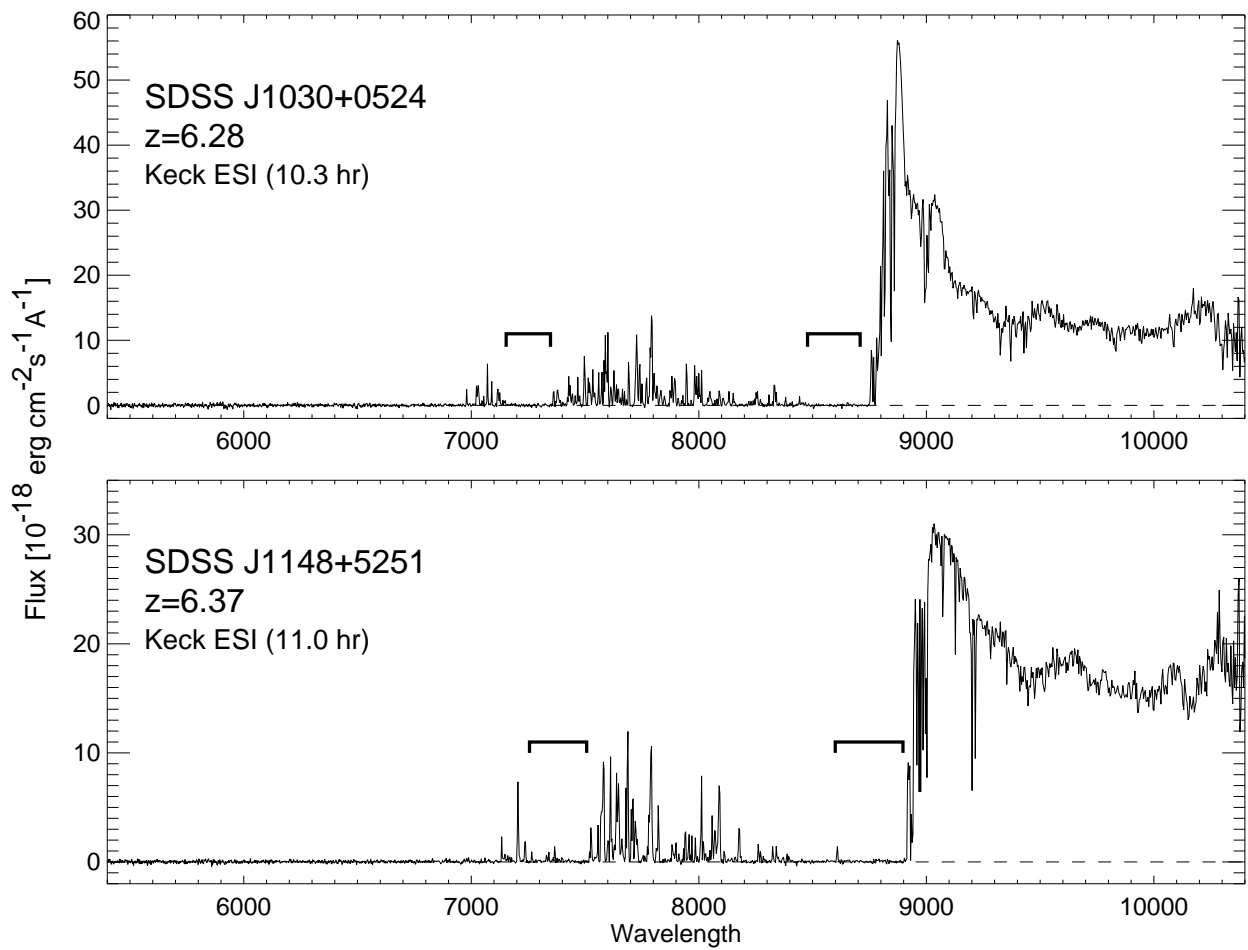}
\caption{
Keck ESI spectra of the two highest redshift quasars. The spectra
have been binned to a resolution $R = \Delta\lambda/\lambda = 2500$
(120~km/s).  The Gunn-Peterson absorption troughs of \lya\ and
\lyb\ are indicated, extending from $z=5.975$ to~6.165 (\qa) and
$z=6.075$ to~6.320 (\qb).
}
\label{fig-spectra}
\end{figure*}

Extraction of spectra from the rectified orders is straightforward.
We use an optimal extraction algorithm with a second-order fit to
the sky along the slit and a spatial profile that is assumed constant for each
order.  For orders where the profile is not detected, the trace position
and width are predicted based on those orders that are detected.
The wavelength scale derived from calibration lamps is adjusted
using the positions of sky lines for each observation.
The final spectra for the quasars are shown in Figure~\ref{fig-spectra}.

\section{Analysis of the Spectra}

High signal-to-noise ratio spectra promise two improvements in our understanding of
the GP troughs.  By reducing the noise in the black regions of the
spectrum, they improve our lower limits on the absorbing optical
depth and may allow us to detect faint light in the dark
troughs.  But since the optical depth limits increase only as the
logarithm of the signal-to-noise ratio, which itself increases only as the square
root of the integration time, it is difficult to make dramatic
improvements in the limits.  The main benefit of high signal-to-noise ratio spectra
is that they allow us to set strong detection limits over narrower
bands in the spectra.  The optical depth distribution is very unlikely
to be a smooth function of redshift; instead, at the transition between
a neutral and an ionized IGM, there will be ionized,
partially transparent bubbles sprinkled along the line of sight.
To detect these bubbles we need to be able to recognize
narrow emission spikes in the trough where the quasar light leaks
through the IGM.  The new spectra presented in the paper are far
better for that purpose than our earlier observations (which were
30~min exposures taken with ESI.)

We begin with an
approach similar to that used by Becker et al.\ (2001).
The residual fluxes in the GP troughs of \lya\ and \lyb\ are
measured directly from the spectra; the fluxes are weighted
using the sky noise from adjacent pixels (as described above
for the extraction) to improve the signal-to-noise ratio while
avoiding bias in the sums.  Then the mean transmission and
optical depth of the IGM are determined by comparing the
residual flux to the (unabsorbed) quasar continuum
at the positions of the troughs.

Because very little neutral hydrogen is required to create a GP
trough in \lya\ (a 1\% neutral fraction will suffice), the presence
of a GP trough due to \lya\ provides only a lower limit to the
amount of neutral hydrogen. The GP trough from the
\lyb\ transition can be a more sensitive test because the
\lyb\ cross-section is a factor of 5.27 lower than \lya.
Interpretation of the \lyb\ trough is complicated by the
fact that it is overlaid by the \lya\ forest at $z\sim5$, which is unpredictable
along a given line of sight due to cosmic variance.  In spite of
that, using fairly secure assumptions, the \lyb\ trough can give
much stronger limits on the neutral hydrogen fraction. We consequently
measure residual fluxes in both the \lya\ and \lyb\ absorption
regions.

We also considered using the \lyg\ absorption since \lyg\ has an
even lower intrinsic optical depth.  However, the \lyg\ GP trough
suffers from overlying absorption by both \lya\ and \lyb\ at lower
redshifts and consequently does not give additional useful information.

It is necessary to estimate the brightness of the quasar continuum
at the \lya\ and \lyb\ troughs in order to measure the transmission
of the IGM.  For this purpose we use the far UV quasar power law
spectrum of Telfer, Zheng, Kriss, \& Davidsen (2002) 
and the composite spectrum from the Large Bright Quasar Survey
(LBQS; Francis et al.\ 1991\footnote{We use the improved
LBQS template created by S.~Morris, described in Brotherton et al.\ (2001)
and available at \url{http://sundog.stsci.edu/first/QSOComposites}.
Note that the LBQS composite is very similar to the FIRST Bright Quasar
Survey composite derived by Brotherton et al.}).
The templates are matched to
the observed spectra at a rest wavelength of $\sim1290$~\AA.  
The Telfer et al.\ power law has a break near 1280~\AA\ and rises less steeply
into the UV than a simple extrapolation of the continuum redward
of \lya.  In the near UV ($\lambda > 1280$~\AA), the spectral
slope is $\alpha_{NUV} = -0.72$ ($f_\nu \propto \nu^\alpha$),
while in the far UV ($\lambda < 1280$~\AA), $\alpha_{EUV} = -1.57$
for radio-quiet objects and $-1.96$ for radio-loud objects.
While there is no evidence that either of our quasars is radio-loud,
we compute continuum levels using both EUV slopes in
order to represent the range of continua seen in quasars.

\begin{figure*}
\plotone{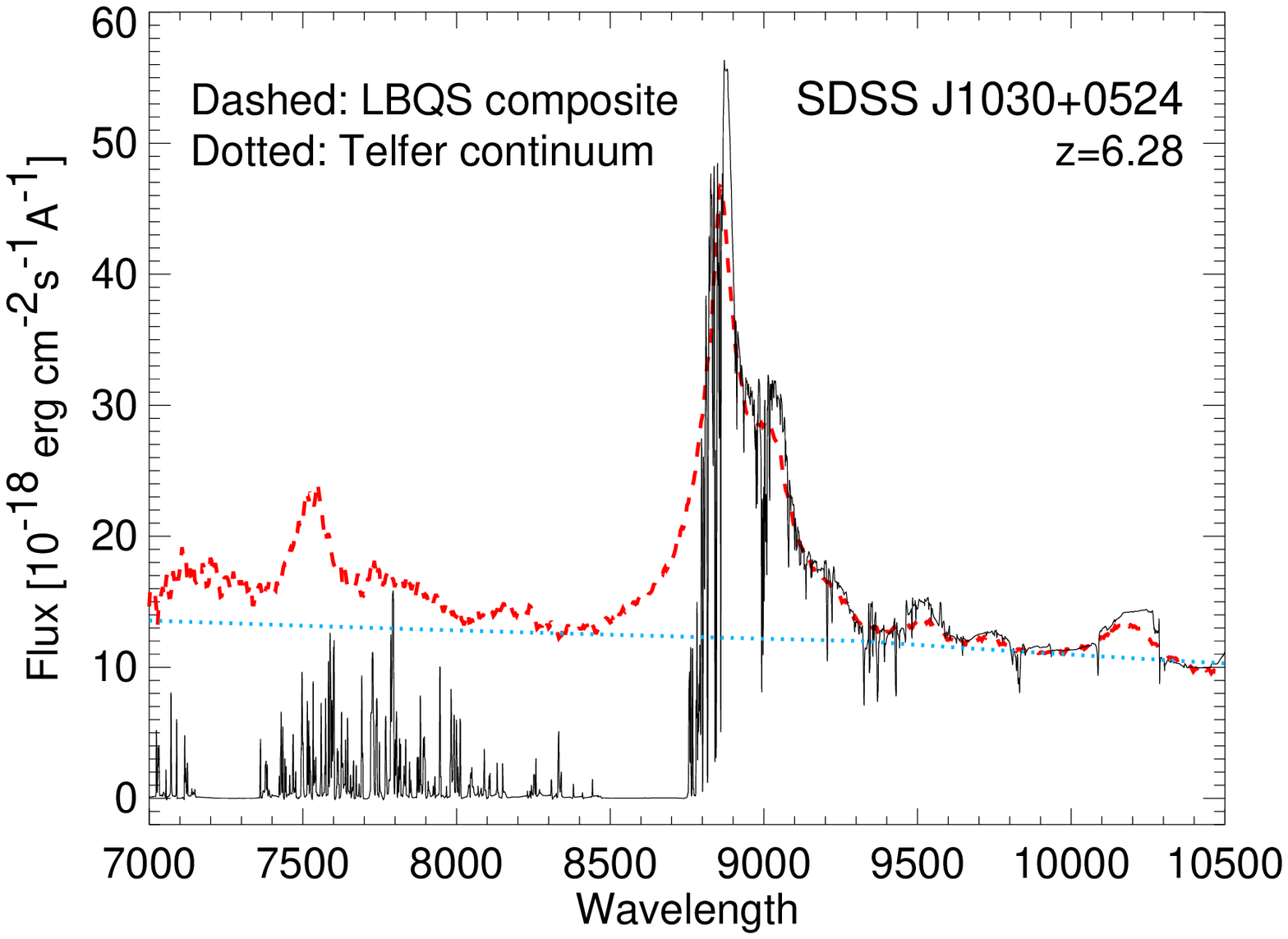}
\caption{
Denoised, full-resolution spectrum of \qa\ with matched templates from
the LBQS and Telfer et al.\ (2002).  The denoising algorithm is
described in the text. The template is a very good match to the quasar
redward of the \lya\ IGM absorption.
}
\label{fig-atemplate}
\end{figure*}

\begin{figure*}
\plotone{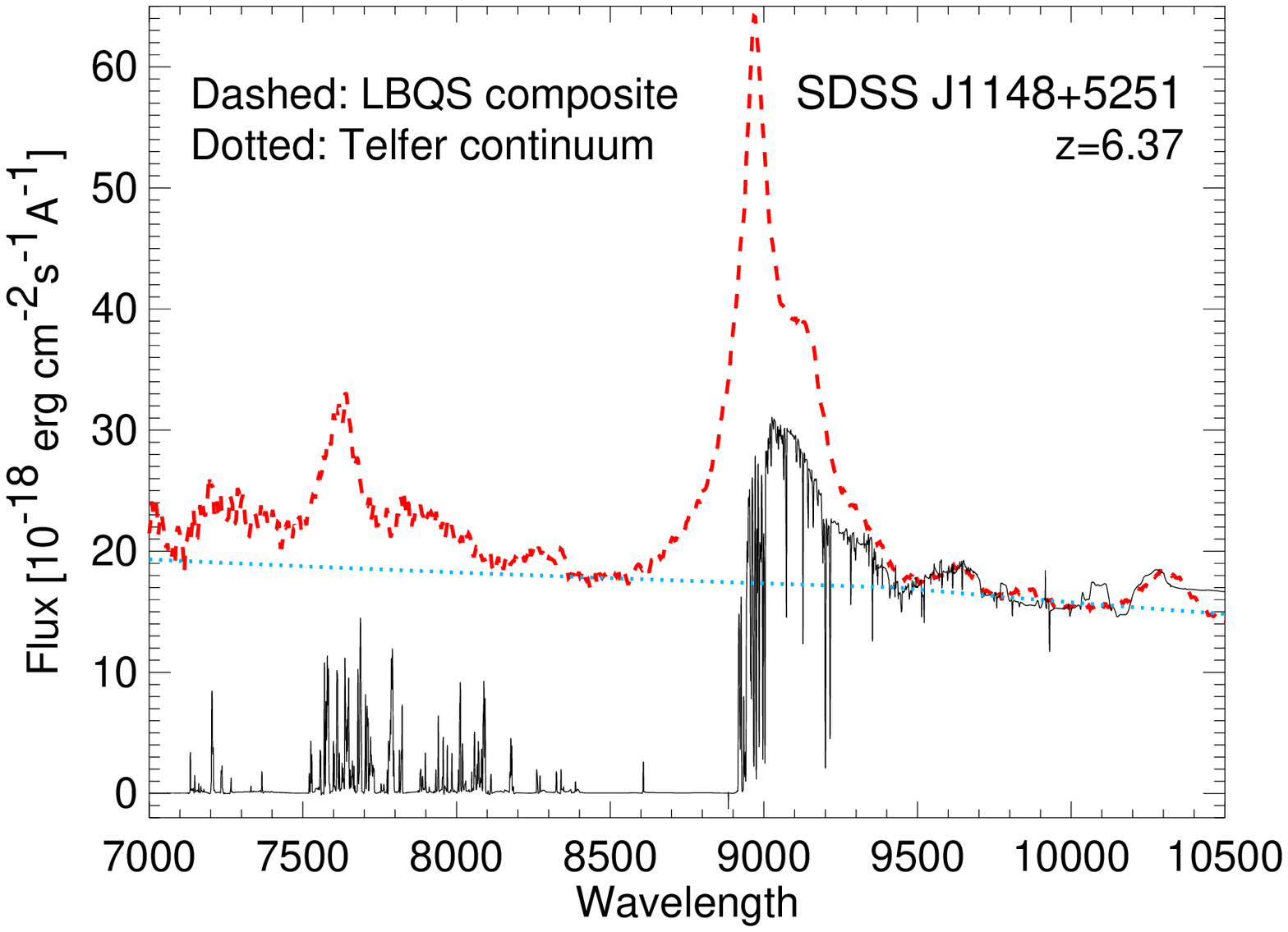}
\caption{
Denoised, full-resolution spectrum of \qb\ with matched templates from
the LBQS and Telfer et al.  The hydrogen emission lines are much weaker
than those in the template.  Our best redshift determination for this
object, $z=6.37$, comes from matching the features redward of
9500~\AA\ with the template.
}
\label{fig-btemplate}
\end{figure*}

Broad emission lines can also contribute to the continuum.  Our spectra
are displayed in Figures~\ref{fig-atemplate} and~\ref{fig-btemplate}
along with the LBQS composite spectrum.  For these figures, our spectra
have been denoised using a wavelet shrinkage algorithm (Donoho \&
Johnstone 1994) that is optimized for the case where the noise
amplitude is variable but known (heteroscedastic data, in the parlance
of statisticians).  We use a Haar (1910) wavelet transform, which
allows error propagation and has the great benefit of keeping the noise
in transform space uncorrelated at each scale before the
filtering is applied.  Cycle-spinning (Coifman
\& Donoho 1995) is utilized to make the result shift-invariant and to
remove the artifacts generated by the square-wave Haar basis
functions.

Denoising thus uses a variable smoothing length to suppress the
noise as much as possible on all scales while remaining consistent
with the spectrum given the estimated noise array.
The resulting spectrum
is heavily smoothed in flat regions (such as the dark troughs),
which suppresses the noise.  (Values in the smoothed
areas are consequently highly correlated.)
The smoothing threshold is
set to $3\sigma$, so only structures more significant than that remain
in the spectrum.  Note that high resolution structures are preserved
(compare the absorption on the blue edge of the \lya\ emission with
Fig.~\ref{fig-spectra}) while the noise in both the continuum and
the troughs is dramatically reduced.

For \qa\ (Fig.~\ref{fig-atemplate}), the LBQS composite is clearly a
good match to the spectrum, so we estimate the continuum in the
\lya\ and \lyb\ absorption troughs using the mean flux in the
relevant wavelength window from the LBQS spectrum.  On the other hand,
\qb\ shows considerably weaker emission lines than the composite
(Fig.~\ref{fig-btemplate}); we therefore use the Telfer et
al.\ power law continuum alone as the estimate of the brightness at the
absorption troughs.  The continua estimated by using these different
templates differ by about 30\%.

\section{Results}

\subsection{\qa}

\begin{figure*}
\plotone{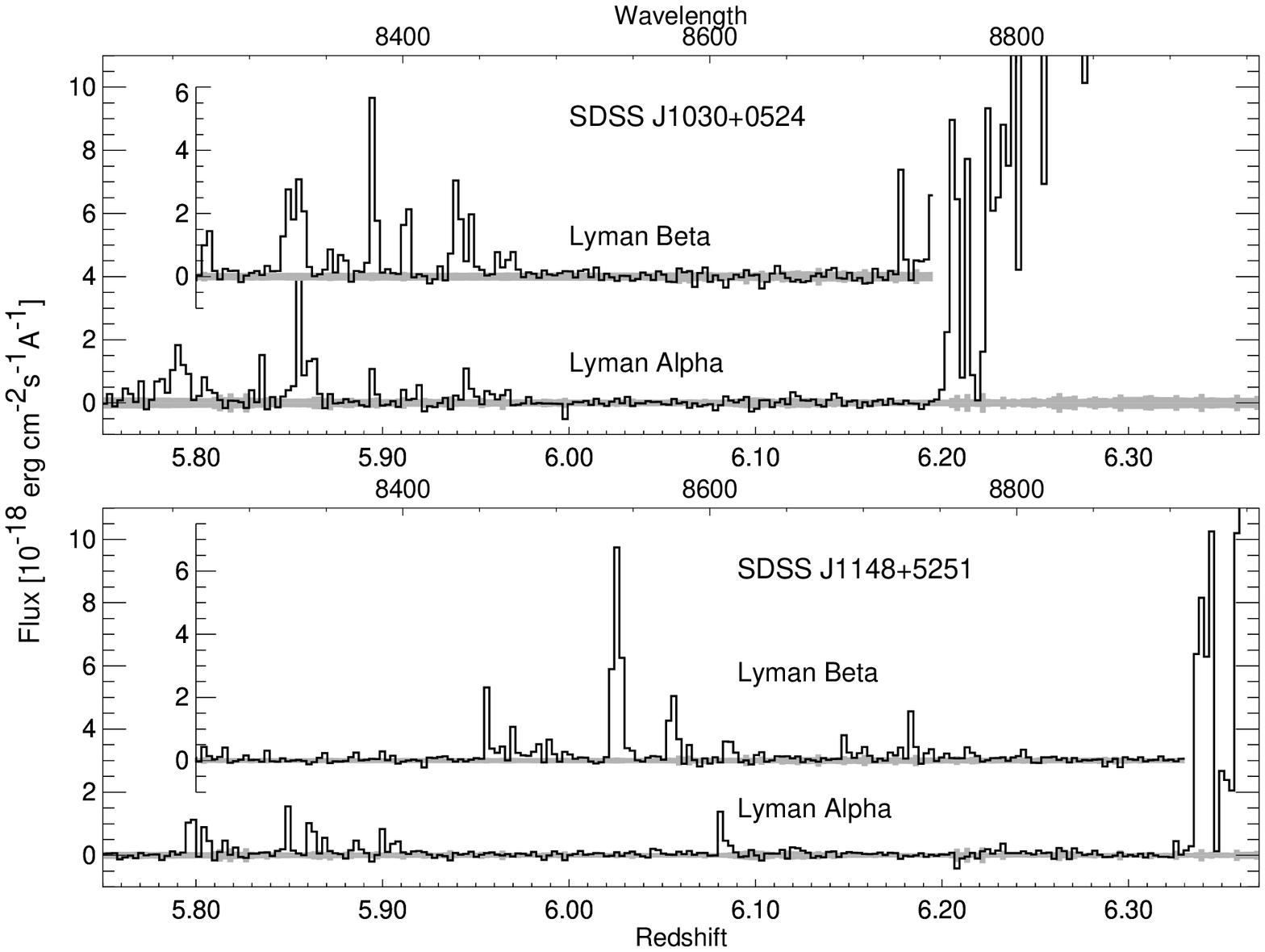}
\caption{
Closeup of the \lya\ and \lyb\ GP troughs, aligned in
redshift, at a resolution of $R=2500$ (120~km/s).  Note that
these spectra have not been denoised, so the pixel values
are independent.  The top axis
shows the wavelength for \lya, and the shaded band shows $1\sigma$
errors.  The absorption is extraordinarily black for \qa, but
\qb\ shows evidence for weak emission in the trough, especially
at \lyb.  The strong peaks in the \lyb\ spectrum
at $z=6.02$ and 6.06 are discussed in
the text; they could be \lya\ emission from a foreground
galaxy at $z=5$.
}
\label{fig-troughs}
\end{figure*}

For \qa, our results can be described very simply: the GP absorption
troughs of both \lya\ and \lyb\ are very black.  There is no evidence
of any residual light in the troughs to the limit of our observations.
The \lya\ and \lyb\ absorption trough details are shown in
Figure~\ref{fig-troughs}.

\begin{deluxetable}{cccccccrrrr}
\tabletypesize{\scriptsize}
\tablecolumns{10}
\tablewidth{0pc}
\tablecaption{Absorption Properties of $z>6$ Quasars}
\tablehead{
\colhead{Object} & \colhead{GP Trough} &
\colhead{$\lambda_{min}$} & \colhead{$\lambda_{max}$} &
\colhead{$z_{min}$} & \colhead{$z_{max}$} &
\colhead{Residual Flux} & \colhead{Continuum} &
\colhead{IGM Transmission} & \colhead{\lya\ Optical} \\
\colhead{} & \colhead{} &
\colhead{(\AA)} & \colhead{(\AA)} & 
\colhead{} & \colhead{} & 
\multicolumn{2}{c}{($10^{-20}\,\hbox{erg}\,\hbox{s}^{-1}\hbox{cm}^{-2}\hbox{\AA}^{-1}$)} &
\colhead{Fraction} & \colhead{Depth} \\
}
\startdata
\qa & \lya & 8510 & 8710 & 6.00 & 6.17 & $1.6 \pm 1.4$ & 1570 & $0.0010 \pm 0.0009$ & $>6.3$\tablenotemark{a} \\
$z_{em}=6.28$
	& \lyb & 7180 & 7349 & 6.00 & 6.17 & $0.8 \pm 1.7$ & 200\tablenotemark{b} & $0.0043 \pm 0.0088$ & $>22.8$\tablenotemark{a} \\
\\
\qb & \lya & 8630 & 8900 & 6.10 & 6.32 & $3.6 \pm 0.8$ & 1730 & $0.0021 \pm 0.0005$ & $6.2\pm0.2$ \\
$z_{em}=6.37$
    & \lyb & 7282 & 7509 & 6.10 & 6.32 & $9.6 \pm 1.0$ & 210\tablenotemark{c} & $0.0465 \pm 0.0047$ & $16.2\pm0.6$ \\
    & \lyb & 7282 & 7509 & 6.10 & 6.32 & $9.6 \pm 1.0$ & 190\tablenotemark{d} & $0.0510 \pm 0.0052$ & $15.7\pm0.6$ \\
    & \lya & 8510 & 8630 & 6.00 & 6.10 & $10.9 \pm 1.1$ & 1740 & $0.0063 \pm 0.0007$ & $5.1\pm0.1$ \\
    & \lyb & 7180 & 7282 & 6.00 & 6.10 & $63.4 \pm 1.4$ & 210\tablenotemark{c} & $0.3020 \pm 0.0067$ & $6.3\pm0.1$ \\
    & \lyb & 7180 & 7282 & 6.00 & 6.10 & $63.4 \pm 1.4$ & 190\tablenotemark{d} & $0.3350 \pm 0.0074$ & $5.8\pm0.1$ \\
\tablenotetext{a}{$1\sigma$ lower limit to the optical depth.}
\tablenotetext{b}{Derived from the LBQS composite spectrum with
a \lya\ forest absorption factor of 0.12.}
\tablenotetext{c}{Derived from the Telfer et al.\ (2002) radio-quiet spectral index, $\alpha_{EUV} = -1.57$
with a \lya\ forest absorption factor of 0.11.}
\tablenotetext{d}{Derived from the Telfer et al.\ (2002) radio-loud spectral index, $\alpha_{EUV} = -1.96$
with a \lya\ forest absorption factor of 0.11.}
\enddata
\end{deluxetable}

The results are summarized in Table~2.  The band limits were chosen
to exclude the first bright peaks on either end of the \lyb\ GP
trough and are identical in redshift for the two troughs.  The
residual fluxes for both the \lya\ and \lyb\ troughs are consistent
with zero.

For the \lyb\ trough, the continuum estimate includes a reduction
factor of 0.12 for the overlying \lya\ forest absorption at $z=5$ (as
determined by Becker et al.\ 2001) The $\taub$ values are converted to
equivalent \lya\ optical depths by multiplying by 5.27, the ratio of
the oscillator strengths of the two lines.  Thus the \lyb\ GP trough
measurements for \qa\ imply $\taua > 22.8$ ($1\sigma$).  The errors
quoted in the table reflect only the statistical uncertainties.  As
found by Becker et al., the strongest limit on the optical depth comes
from the \lyb\ trough, despite the additional uncertainty that comes
from the overlying \lya\ absorption.

These results are consistent with the measurements of Becker et
al.\ (2001) (confirmed by Pentericci et al.\ 2002) but are naturally
stronger given the much improved signal-to-noise ratio of the data.
Our limits on the residual flux in the GP troughs and the \lya\ optical
depth of the IGM are lower by the expected factors given the longer
integrations reported here.  The limits remain consistent with zero
flux in the GP troughs of \qa.

\begin{figure*}
\plotone{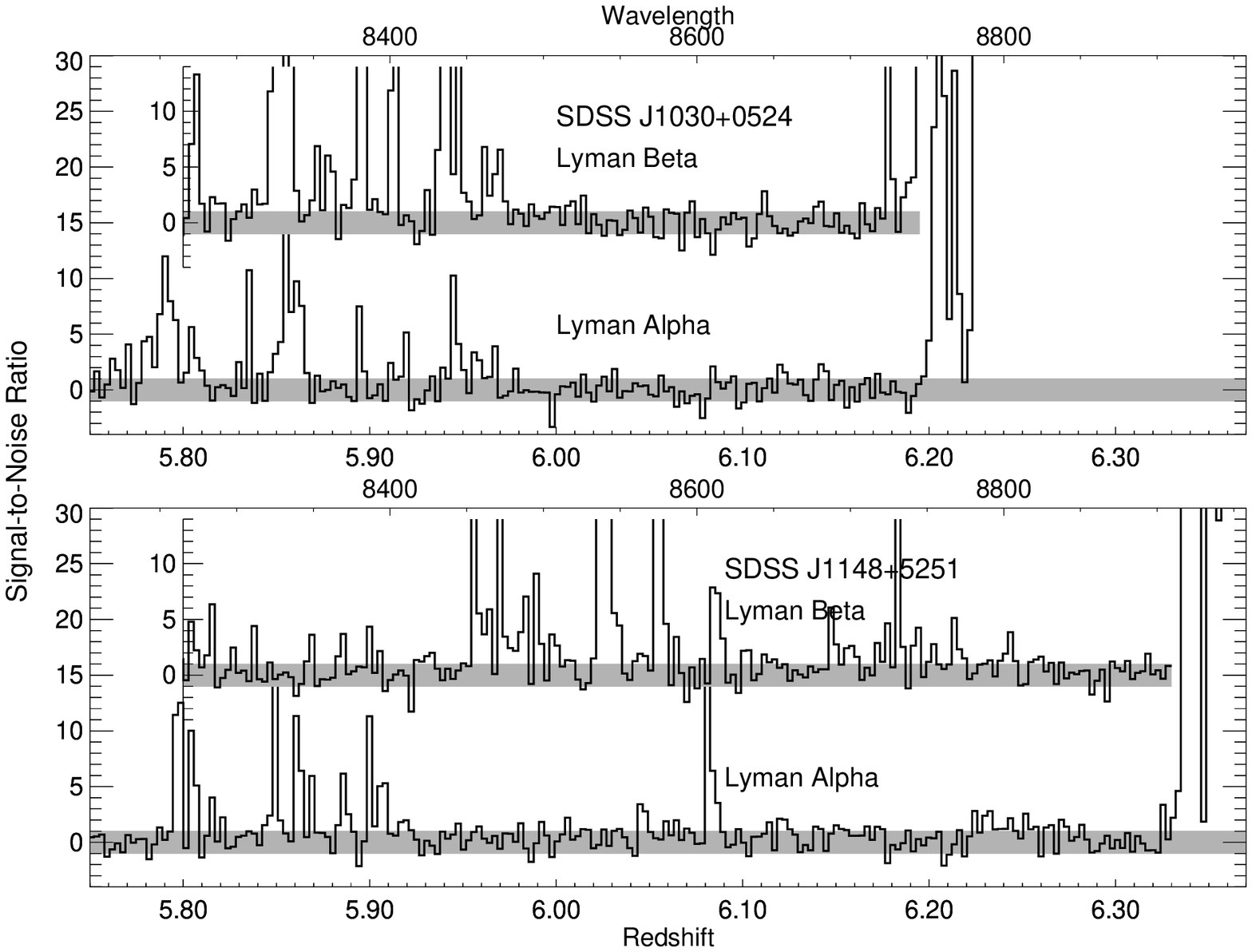}
\caption{
Signal-to-noise ratios for the \lya\ and \lyb\ absorption troughs in
\qa\ and \qb.
}
\label{fig-troughsnr}
\end{figure*}

There is no evidence for any significant peaks in either trough.
Figure~\ref{fig-troughsnr} displays the signal-to-noise ratio
of the spectra in the troughs.  The distribution is consistent
with zero-mean noise in the troughs; the reduced $\chi_\nu^2$ for the
combined \lya\ and \lyb\ troughs is 0.97.

\subsection{\qb}

The GP troughs for \qb\ are also shown in Figures~\ref{fig-troughs}
and~\ref{fig-troughsnr}.  The mean
transmission properties of the troughs are given in Table~2, which
includes the transmission measured for both a high redshift range
($6.1<z<6.32$) and a lower redshift interval ($6.0<z<6.1$).  Note that
we now quote a measurement, not a lower limit, on $\tau$, as flux has
been detected in the troughs.  The correction factor for \lya\ forest
absorption overlying the \lyb\ GP trough was taken to be 0.11, from
Figure~2 of Becker et al.\ (2001).  The $1\sigma$ errors on the
transmission are smaller than for \qa, both because the seeing and
photometric conditions were slightly better for \qb\ and because its
continuum is brighter.

\begin{figure*}
\plotone{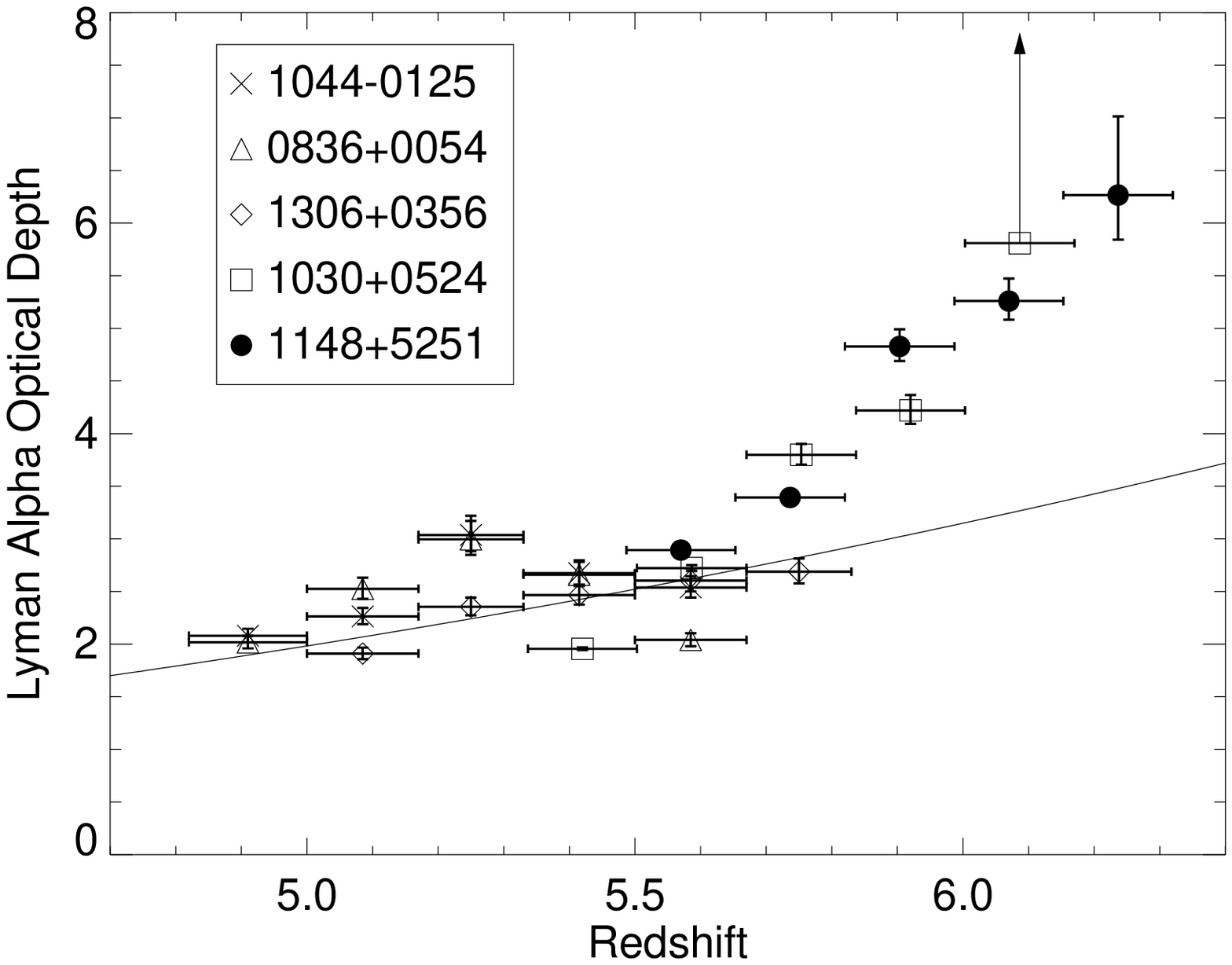}
\caption{
\lya\ optical depth as a function of redshift from high-redshift
quasars.  Error bars (and the lower limit for \qa) are $2\sigma$.  Data
for \qa\ and \qb\ are from this paper; other data are from Fan et
al.\ (2001).  The line shows the expected evolution when the density of
\lya\ forest clouds scales as $(1+z)^{5/2}$ (also from Fan et al.) The
implied $\taua$ values from the \lyb\ GP trough transmission are
substantially stronger:  $\tau>22$ (\qa) and $\tau=16$ (\qb).  We
believe that highest redshift bins for \qb\ are probably contaminated
by light from an intervening galaxy and so should be considered lower
limits.
}
\label{fig-tau_vs_z}
\end{figure*}

In contrast to \qa, there is light detected in the GP troughs of
\qb\ both in narrow spikes and in broader bands.  The mean transmission
nonetheless remains low.  Figure~\ref{fig-tau_vs_z} summarizes the
\lya\ optical depth as a function of redshift, including additional
high-redshift quasars from Fan et al.\ (2001).  There is a noticeable
steepening of the $\tau(z)$ relationship at $z\sim5.8$.

A closer examination of the troughs reveals some interesting features.
The \lya\ trough extends from $z=5.93$ to 6.32 and looks dark at
high resolution except for a weak emission spike at $z=6.08$.  The
reality of that peak is supported by its coincidence with a
\lyb\ emission peak at the same redshift (Fig.~\ref{fig-troughs}).  We
believe this is likely to be a real detection of an ionized bubble in
the IGM that permits quasar light to leak through at the redshift of
the bubble; this feature is discussed in detail below (\S\ref{section-6.08}).

Several emission spikes are detected in the \lyb\ trough as well.  Two
especially strong peaks are seen at
7205~\AA\ and 7236~\AA\ (\lyb\ redshifts of $z=6.03$ and 6.06).
There also is
significant non-zero flux in the troughs even between the peaks (best
seen in Fig.~\ref{fig-troughsnr}).  If this flux is quasar light
transmitted by the IGM, it certainly suggests strongly that the IGM is
not very neutral at $z\sim6$.  That interpretation is not so clear cut,
however. We argue below (\S\ref{section-intervening})
for the alternative interpretation that both
the \lya\ and \lyb\ troughs are contaminated by light from an
intervening galaxy at $z=5$.
If that suggestion is correct, the
optical depths derived from the troughs of \qb\ are only lower limits.

\section{Discussion}

\subsection{The $z=6.08$ emission feature in \qb}
\label{section-6.08}

The universe at $z\sim6$ was very likely ionized by galaxies rather
than by quasars, since there do not appear to be enough
quasars to do the job (Fan et al.~2001). There must have been regions in the preionized
universe where a concentration of unusually luminous galaxies
(possibly the precursor to a massive cluster) was capable of ionizing
the nearby IGM, creating an ionized bubble.  It is therefore not
surprising to find weak emission spikes embedded in the mostly
black GP troughs (Figs.~\ref{fig-troughs} and~\ref{fig-troughsnr})
where the line-of-sight to the quasar passes through small ``holes''
in the IGM \lya\ and \lyb\ transmission created by ionized bubbles.
That can also explain the detection of \lya-emitting galaxies at
$z\sim6.5$ (Hu et al.\ 2002, Kodaira et al.\ 2003), which in this
picture ought to be found in concentrations with other galaxies.

According to the scenario presented by Cen (2003), the universe is
first ionized at $z\sim15$ by Pop~III stars and then recombines to
become substantially neutral, with neutral fractions remaining
above 0.1 until the second reionization occurs at $z\sim6$.  At
$z=6$ to 7 the neutral fraction $\nhrat$ is typically $\sim
0.15$.  As Cen points out, that makes it easier to detect high
redshift galaxies; it also makes it easier for lower luminosity
sources to create ionized bubbles.

The peak seen in the \lya\ trough of \qb\ at 8600~\AA\ ($z=6.08$)
has an intensity about 8\% of the extrapolated (pre-absorption)
continuum, implying an IGM optical
depth of $\taua=2.5$.  The optical depth of this window
at \lyb\ would be only about 0.5, which would produce a very strong
peak in the \lyb\ trough; but since the overlying \lya\ forest
absorption is expected to reduce that by a factor of 0.11 (corresponding
to an overlying optical depth $\tau=2.2$), the
expected amplitude of the \lyb\ trough peak is in fact very similar
to that observed, albeit with a rather large uncertainty in
the \lya\ forest attenuation.

How large would an ionized region need to be to produce this feature?
The damping wings from neutral hydrogen on either side of the ionized
bubble create a high optical depth through the region unless the bubble
is relatively large.  This led Barkana (2002) to question Djorgovski
et al.'s (2001) argument that dark regions in the spectrum of a lower redshift
were created by neutral regions of the IGM.  Miralda-Escud\'e (1998;
see also Barkana 2002) derived the \lya\ damping wing optical depth for
hydrogen extending from $z_1$ to $z_2$ when observed at a wavelength
$\lambda_\alpha(1+z)$:
\begin{equation}
\tau_{\rm damp} =  6.43\times10^{-9}
\, \tau_{\rm GP}
\, \left( \nhratover \right)
\times \left[
I\left(1 + z_2 \over 1 + z\right) -
I\left(1 + z_1 \over 1 + z\right) 
\right] \quad,
\label{eqn-taudamp}
\end{equation}
where $\tau_{\rm GP}$ is the Gunn-Peterson optical depth,
\begin{equation}
\tau_{\rm GP} = 3.42\times10^5
\left(1+z \over 7.08\right)^{3/2}
\quad ,
\label{eqn-taugp}
\end{equation}
and we are using a $\Lambda$ cosmology with
$H_0 = 70$, $\Omega_\Lambda = 0.7$, $\Omega_m = 0.3$, and $\Omega_b = 0.04$.
The function $I(x)$ is defined to be
\begin{equation}
I(x) = x^{1/2} \left( {x^4 \over 1-x} + {9 \over 7}x^3 +
{9 \over 5}x^2 + 3x + 9\right) -
{9 \over 2} \ln \left| 1 + x^{1/2} \over 1 - x^{1/2} \right|
\quad .
\end{equation}
Given an ionized region at redshift $z$, Eqn.~(\ref{eqn-taudamp}) allows us
compute the combined damping wing optical depth from the neutral
regions on either side of the bubble.  Although the constant factor
multiplying $\tau_{\rm GP}$ in Eqn.~(\ref{eqn-taudamp}) is small,
the leading term in $I(x)$ can be very large near $z=z_1$ or $z_2$,
so $\tau_{\rm damp}$ can be substantial.

\begin{figure*}
\plotone{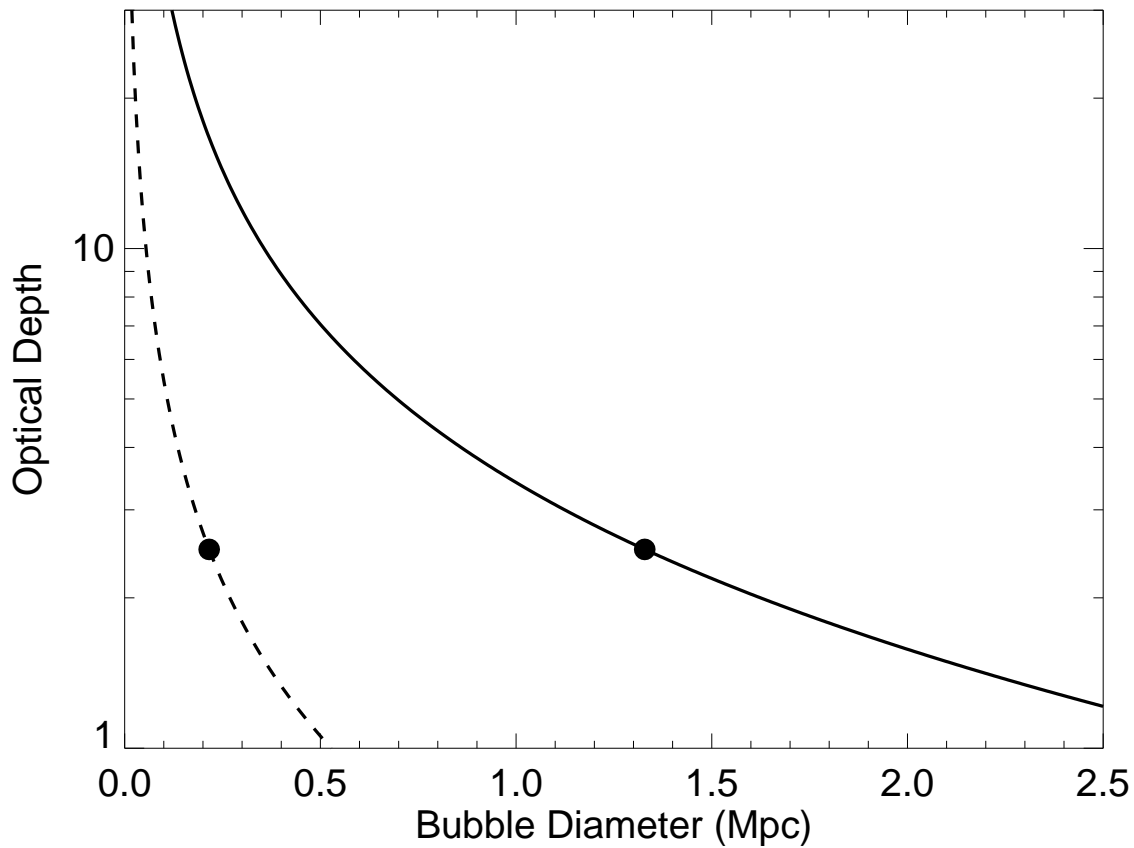}
\caption{
Optical depth from damping wings of \lya\ as a function of diameter for
an ionized bubble at $z=6.08$.  For small
bubbles in a completely neutral IGM (solid line),
the damping wings strongly suppress the transmission.
If the IGM is already partly ionized ($\nhrat = 0.15$; dashed
line) then the optical depths are smaller by the same factor.
The dots show the bubble diameter required to produce the
observed peak in the \lya\ trough of \qb\
(1.33~Mpc for a neutral IGM or 0.22 Mpc for a partially
ionized IGM.)
}
\label{fig-taudamp}
\end{figure*}

Figure~\ref{fig-taudamp} displays the resulting optical depth for a
bubble located at $z=6.08$.  The neutral IGM is assumed to extend from
$z=6$ to $z=6.32$, but the results are relatively insensitive to those
limits.  The bubble diameter required to produce a window with
$\tau=2.5$ is 1.33~Mpc if the IGM is completely neutral outside the
bubble.  If the IGM is already partly ionized, the required bubble size
is much smaller (0.22~Mpc if $\nhrat=0.15$).

Barkana (2002) computed the diameter of an \ion{H}{2} bubble
as a function of the halo mass $M$ of the ionizing
source (including the escape probability for ionizing
photons in a low metallicity galaxy) to be
\begin{equation}
D_S = 0.150 \, {\rm Mpc} \,
\left(7.08 \over 1+z \right)
\left(
{N_{\rm ion} \over 20}
{M \over 10^9\,{\rm M}_\sun}
\right)^{1/3} \quad ,
\label{eqn-ds_vs_m}
\end{equation}
where $N_{\rm ion}$ is the number of ionizing photons per baryon that
escape the galaxy.  We can invert this to get the mass required to
produce a bubble of a given diameter:
\begin{equation}
M = 7\times10^{11}\, {\rm M}_\sun
\left(
{ D_S \over 1.33\,{\rm Mpc}}
{ 1+z \over 7.08 }
\right)^3
{20 \over N_{\rm ion}} \quad .
\label{eqn-ionmass}
\end{equation}
We conclude that a plausible star-forming galaxy (or group of galaxies)
along the line of sight can produce the observed peak at $z=6.08$ in
the \lya\ and \lyb\ GP troughs of \qb, even if the surrounding IGM is
completely neutral.  Obviously if the IGM is partly ionized, the
required bubble diameter would be much smaller (Fig.~\ref{fig-taudamp})
and so the required galaxy mass would also be far smaller
($<10^9\,{\rm M}_{\sun}$).  With future
instrumentation it will be interesting to search the vicinity of the
quasar for evidence of star-forming galaxies at $z=6.08$ that are the
source of the ionizing continuum radiation required to make this
bubble.

\subsection{Evidence for an intervening $z=5$ galaxy}
\label{section-intervening}

The strong spikes seen in the \lyb\ trough of \qb\ at 7205~\AA\ and
7236~\AA\ are more difficult to explain as the result of ionizing
bubbles in the IGM.  The strength of these peaks is not easily
reconciled with the absence of corresponding \lya\ peaks, given the
expected overlying \lya\ forest absorption.  The peak at 7205~\AA\ has
an amplitude $8.7\pm0.3$
($\times10^{-18}\,\hbox{erg}\,\hbox{s}^{-1}\hbox{cm}^{-2}\hbox{\AA}^{-1}$),
while the brightest plausible continuum at that wavelength is 24 (from
the LBQS template spectrum in Fig.~\ref{fig-btemplate}).  If we assume
that the 7205~\AA\ peak is quasar light leaking through the $z{=}6$
\lyb\ and $z{=}5$ \lya\ absorption, the implied combined optical depth is
$\tau(\hbox{\lya@}z{=}5) + \tau(\hbox{\lyb@}z{=}6) = 1.0$.  But there is no
hint of a \lya\ peak at 8539~\AA:  the flux limit there is $0.1\pm0.2$
with an estimated continuum level of 17, implying $3\sigma$ upper
limits $\tau(\hbox{\lya@}z{=}6)>3.3$ and $\tau(\hbox{\lyb@}z{=}6) =
\tau(\hbox{\lya@}z{=}6) / 5.27 > 0.6$.  The overlying \lya\ forest
absorption must then be $\tau(\hbox{\lya@}z{=}5) < 0.4$.

We conclude that if the 7205~\AA\ peak is the result of an ionized
bubble at $z=6.03$, it must fall in a highly transparent window in the
$z=5$ \lya\ forest that barely attenuates the continuum at all,
allowing $>65$\% of the light through.  High resolution spectra of the
\lya\ forest at $z\sim5$ (Songaila, Hu, Cowie \& McMahon 1999;
Djorgovski et al.\ 2001; Songaila \& Cowie 2002) do show small
transparent windows, but they are rare.  The spectra presented in this
paper are typical.  There is one transparent window seen at
7800~\AA\ ($z=5.4$) in our spectrum of \qa\ (Fig.~\ref{fig-atemplate}),
and no windows at all are seen in \qb.  In the $5<z<5.5$ \lya\ forest,
only 0.5\% of the bandpass has optical depths less than 0.4.  Songaila
\& Cowie (2002) discuss the distribution of the transmitted fraction as
a function of redshift; using their analytical model, at $z=5$ only
1.7\% of the spectrum is expected to have a transmission $>0.65$.  This
fraction drops to only 0.6\% if we use the $1\sigma$ upper limit,
$\tau(\hbox{\lya@}z{=}5) < 0.2$.

We conclude that it is problematic to reconcile the strong peak at
$z\sim6.03$ in the \lyb\ trough with the absence of a corresponding
\lya\ peak.  Unless the \lya\ forest is extraordinarily transparent at
the position of the peak, we would expect a strong, easily detected
peak in the \lya\ GP trough.

\begin{figure*}
\plotone{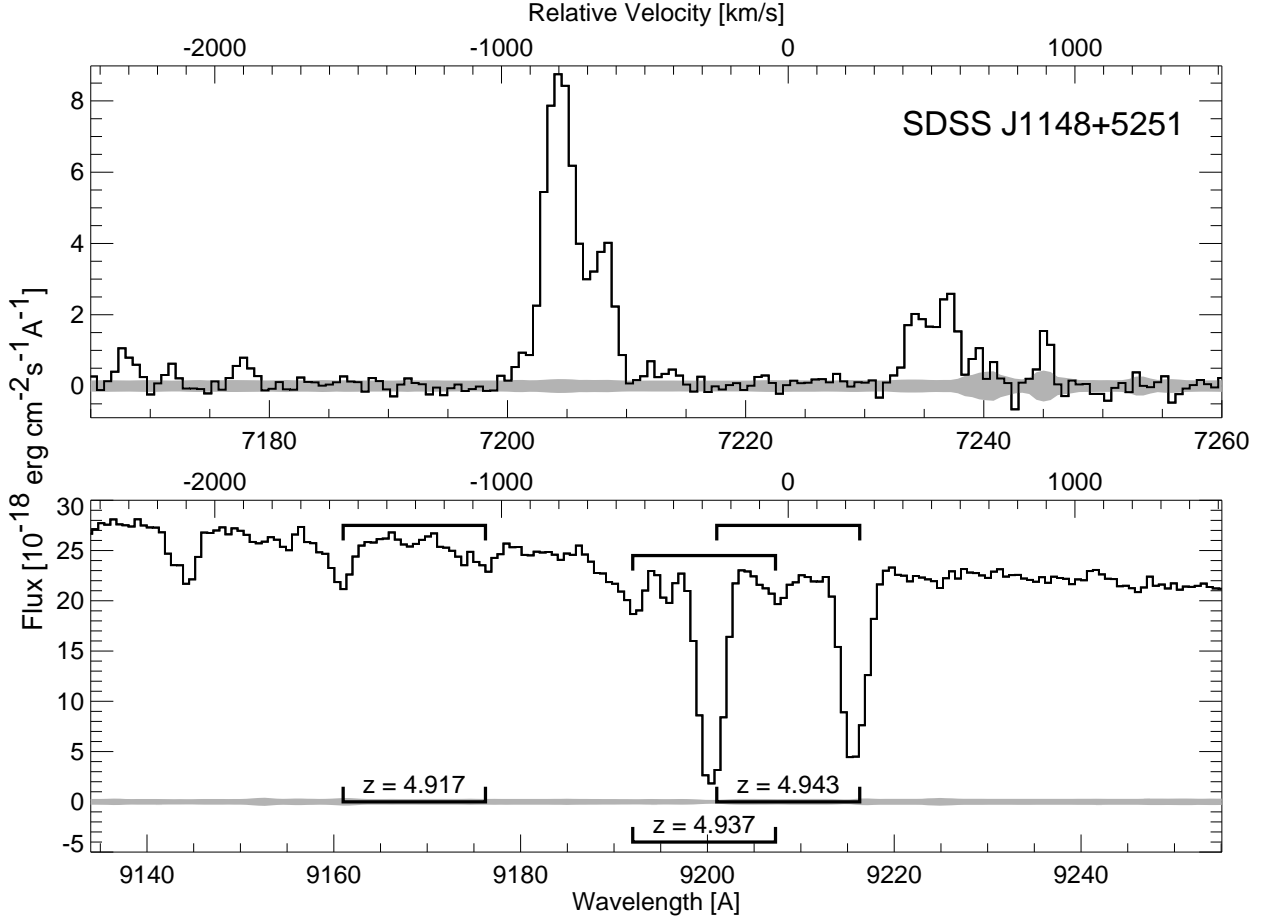}
\caption{
Comparison of emission in the \lyb\ GP trough of \qb\ with absorption by
the \ion{C}{4} 1548.20, 1550.77 doublet, under the assumption that the
emission lines are \lya\ from an intervening galaxy at the same
redshift as the \ion{C}{4} absorption, $z=4.943$.
Probable \ion{C}{4} absorption systems (having both the
correct separation and the correct optical depth ratio) are marked by
brackets in the lower panel.  The panels are aligned
in redshift so that the \lya\ peak would fall midway between
the \ion{C}{4} doublet components for perfect alignment.
The velocity axis is registered to the strongest
\ion{C}{4} at $z=4.943$.  The data are shown at the full
resolution of ESI (a substantially higher resolution than
is shown in Fig.~4.)
}
\label{fig-civ}
\end{figure*}

There is, however, another possible explanation for the strong
peaks in the \lyb\ trough of \qb.  A careful examination of the
quasar spectrum (Fig.~\ref{fig-civ}) reveals a strong \ion{C}{4} absorption
feature at 9200A ($z=4.943$). That suggests a second interpretation
for the spikes, namely that they are \lya\ emission from a foreground galaxy
at $z=4.943$. Usually one does not see emission associated with \ion{C}{4}
absorbers; but ordinarily such emission lines would be lost in the
\lya\ forest and not noticed.  This particular emission line,
falling as it does in the \lyb\ trough, stands out like a sore
thumb.  Moreover there are multiple \ion{C}{4} absorption systems
and multiple (putative) \lya\ emission components seen, suggesting
the presence of a fairly massive mass concentration.

Is it plausible that the emission in the \lyb\ trough could be from
an intervening $z=4.9$ galaxy?  The implied \lya\ luminosities are
similar to those of \lya\ emitters detected in the Subaru deep
field at $z=4.86$ (Ouchi et al. 2003).  Moreover, the velocity
offset between the \lya\ emission and the \ion{C}{4} absorption
seen in Fig.~\ref{fig-civ} is in the range typically
observed for systems that show both \lya\ emission and metal line
absorption ($\sim500$~km/s, Shapley et al. 2002).

The main argument against the intervening emission model is that
\lya\ emitters are sufficiently rare on the sky that it is unlikely
to find one along a random line-of-sight.  But perhaps this is not
a random line-of-sight.  Figure~\ref{fig-civ} shows that
there are both multiple \lya\ emission components and multiple \ion{C}{4}
absorption components present in the spectrum (note particularly
the structure in both components of the strong \ion{C}{4} doublet
at 9200~\AA.)
The two presumed \lya\ lines are split by 1200~km/s, and the \ion{C}{4}
structure extends over at least 250~km/s.  It is possible that
there is a massive structure (a protocluster?) along the
line-of-sight that is amplifying the quasar through gravitational
lensing.  If that is correct, then the likelihood of finding a
\lya\ emitter in front of the $z=6.37$ quasar is in fact not so
small, since the foreground object assisted in the discovery of
the quasar by boosting its apparent brightness.
Even lensing
by a factor of two might create a strong discovery bias since
the luminosity function is steep for bright quasars
(Wyithe \& Loeb 2002; Comerford, Haiman \& Schaye 2002).

\begin{figure*}
\epsscale{0.7}
\plotone{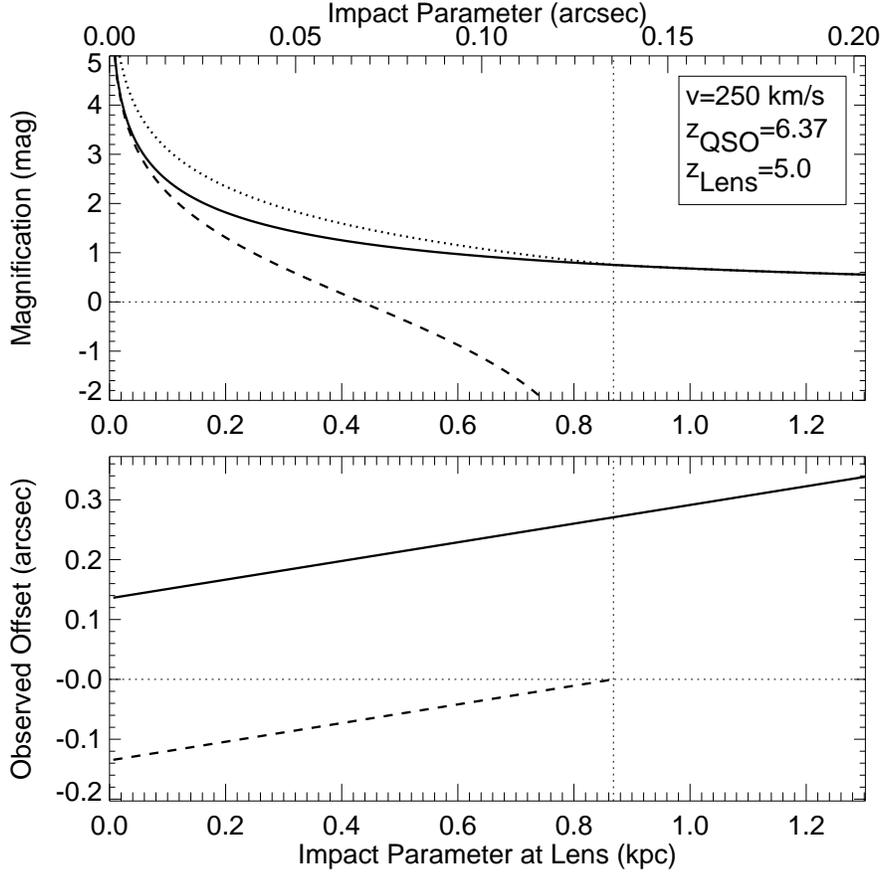}
\epsscale{1.0}
\caption{
Gravitational lensing magnification (top) and splitting (bottom) of
multiple components for a singular isothermal sphere with a velocity
dispersion of 250 km/s.  The solid line shows the primary, brighter
component; the fainter second image that appears for impact parameters
less than the Einstein radius is shown with a dashed line.  The total
magnification when the two components are combined is shown with a
dotted line.  Note that the Einstein ring radius is quite small for a
quasar at $z=6.4$ being lensed by a galaxy at $z=5.0$, so even high
magnification factors are not ruled out by the absence of multiple
components in the image.
}
\label{fig-lens}
\end{figure*}

For the more familiar case where a nearby ($z<1$) galaxy is lensing
a high-redshift quasar, the lensing magnification is limited to
less than $\sim2$ because stronger lensing produces multiple images
with a noticeably wide splitting.  But when the lens redshift is
$z\sim5$, as in the scenario proposed here, the Einstein ring radius
is much smaller and so the splitting could not have been detected
from the ground (in agreement with the imaging observations by
Fan et al.\ 2003.)
Figure~\ref{fig-lens} shows the magnification and observed
image positions as a function of
impact parameter for a singular isothermal sphere galaxy model with
a velocity dispersion of 250~km/s.  This velocity dispersion is consistent
with the velocity structure seen in the strong \ion{C}{4} absorption
system and is slightly larger than the dispersion expected for an $L_*$ galaxy
($v=220$~km/s).  In this geometry, the multiple image separation
is only 0.3~arcsec, which would be difficult to detect even if the
components were of equal brightness.

Note also that since there are multiple \ion{C}{4} absorption
and \lya\ emission components, it is
possible that there are multiple lensing galaxies along the line-of-sight
to this source.  That could possibly increase the amplification
factor further, although it would also lead to larger splitting
and shifts.
We coadded the two-dimensional spectra of \qb\ for the Echelle
order that includes the \lyb\ trough to search for evidence that
the peaks are either broader than the quasar trace or are shifted relative
to the center of the trace.  Neither effect was seen.  The 
width of the peaks agrees with the width of the rest of
the order, and the central positions agree to better than 0.1\arcsec\
(which is the approximate uncertainty in the trace position.)

This appears to be the most problematic aspect of interpreting these
peaks as \lya\ emission from a foreground lensing galaxy: we see no
direct evidence any any images (including the spectrum) for lensing.
That does not contradict the lensing hypothesis given the
small lens scale at $z=5$, but neither does it provide any additional
support.  And there is certainly reason to think it unlikely that we
would stumble across a lens with such a small Einstein ring radius,
since the area on the sky that is lensed by such objects is rather
small.  But as we discuss below, the additional supporting evidence in
favor of the hypothesis is sufficient to make a reasonably strong
case.

\subsection{Continuum emission in the troughs of \qb}
\label{section-continuum}

A puzzling aspect of the \lya\ and \lyb\ GP trough transmissions
for \qb\ is that the \lya\ residual flux is far higher than expected
from the \lyb\ flux.  This is reflected in Table~2 by the fact that
$\taua$ inferred from the \lyb\ trough is far higher than the
optical depth actually measured in \lya.  Either there is too much
light in \lya\ trough or there is too little in the \lyb\ trough
(which presumably would require a great deal of extra absorption in
the $z\sim5$ \lya\ forest.) Neither explanation appears particularly
palatable based on a conventional model of a quasar with IGM
absorption.

An intervening \lya-emitting galaxy at $z=5$ offers a more natural
explanation: the \lya\ and \lyb\ troughs are contaminated by
continuum emission from the galaxy.  We find that the observed
brightness of the continuum is fully consistent with the hypothesis
of an intervening galaxy, which strongly supports our suggestion.
Wyithe \& Loeb (2002) pointed out that a lensing galaxy could
contaminate the black GP trough, although they were concerned mainly
with lensing by low redshift galaxies (which is a priori considerably
more likely.)

Between 8420~\AA\ and 8900~\AA\ ($z=5.93$ to 6.32), the only obvious
emission in the \lya\ trough is the peak at $z=6.08$ discussed
above (Fig.~\ref{fig-troughs}).  If we sum the light in that
wavelength range, excluding the peak ($8600<\lambda<8620$), we get
an estimate for the continuum level of $3.9 \pm 0.6\times10^{-20}
\,\hbox{erg}\,\hbox{s}^{-1}\hbox{cm}^{-2}\hbox{\AA}^{-1}$.  We take
that as an estimate of the brightness of the galaxy continuum.
The AB magnitude of the flux detected in the \lya\ trough is
26.4, which corresponds to an absolute AB magnitude of $-21.9$
at a rest wavelength of 1450~\AA\ if the light comes from a
galaxy at $z=5$.

The observed equivalent width (EW) of the peak at 7205~\AA\ is then
930~\AA, and the rest-frame EW is 160.  This EW is perfectly consistent
with the EW distribution determined by Malhotra \& Rhoads (2002) for
\lya-emitting galaxies at $z=4.5$.  It is slightly lower than the
median EW for their sample, as is expected for a more massive galaxy
(Malhotra \& Rhoads, private communication).

\subsection{The extent of the Str\"omgren sphere}
\label{section-redshift}

Many authors have discussed the possibility that high-redshift
quasars are selected because they are strongly lensed.  Wyithe \&
Loeb (2002) and Comerford, Haiman, \& Schaye (2002) determined how
lensing might bias the quasar luminosity function.
Haiman \& Cen (2002; hereafter HC) considered the
specific case of lensing of \qa; they argue that if the quasar is
embedded in a largely neutral IGM, the large extent of its \ion{H}{2}
region implies that it must in fact have a large ionizing luminosity
and so cannot have been too strongly lensed.  If the ionizing
luminosity were too low, the IGM near the quasar would have remained
neutral and so the \lya\ emission line would have been much more
heavily absorbed on the blue side.  This argument also sets a
minimum age for the quasar, since it must have been in existence
long enough to ionize a substantial Str\"omgren sphere.

The HC result is directly relevant to the question of whether
\qb\ could be strongly gravitationally lensed.  On the blue side
of \lya, emission is detectable to $z=6.32$ (Fig.~\ref{fig-troughs}),
a comoving distance of 2.8~Mpc from the quasar.  (Note that this
distance is somewhat uncertain due to the difficulty in determining an
accurate redshift for the quasar; this is discussed further
below.) According to HC --- see also Madau \& Rees (2000), Cen \&
Haiman (2000) and references therein --- the radius of a Str\"omgren
sphere ionized by a high-redshift quasar as a function of time is
approximately
\begin{equation}
R_S = 7.0\,\hbox{Mpc} \, \left(
 {{\dot N}_{ph} \over 10^{58}\,\hbox{s}^{-1}} \,
 {t_Q \over 10^{7}\,\hbox{yr}}
\right)^{1/3} \,
\left(1+z_Q \over 7.37\right)^{-1} \quad ,
\label{eqn-rs1}
\end{equation}
where ${\dot N}_{ph}$ is the ionizing photon luminosity, $t_Q$ is
the age of the quasar, and as above we use cosmological parameters
$H_0 = 70$, $\Omega_\Lambda = 0.7$, $\Omega_m = 0.3$, and $\Omega_b
= 0.04$.  This formula results from matching the total number of
ionizing photons emitted to the number of hydrogen atoms within a
sphere, and it ignores both recombinations (which reduce the radius)
and the Hubble expansion (which increases it.)  It uses the mean
IGM density assuming that the IGM contains nearly all the baryons;
if the universe is locally overdense by a factor $D$, the radius
is reduced by $D^{-1/3}$.

Inverting Eqn.~(\ref{eqn-rs1}) to solve for the quasar parameters,
we find
\begin{equation}
{{\dot N}_{ph} \over 10^{58}\,\hbox{s}^{-1}} \,
 {t_Q \over 10^{7}\,\hbox{yr}} = 0.063
\left(
  { R_S \over 2.8\,\hbox{Mpc} }
  { 1+z_Q \over 7.37 }
\right)^3 \quad .
\label{eqn-ndott}
\end{equation}
The ionizing luminosity for \qb\ (assuming no lensing) is rather
uncertain.  We have estimated it using two different quasar spectral
templates.  The Telfer et al.\ (2002) quasar template, which we have used
for the \lya\ and \lyb\ continuum normalization, gives an ionizing
photon luminosity of
${\dot N}_{ph} = 0.9\hbox{--}1.3\times10^{58}\,\hbox{s}^{-1}$,
where the smaller value assumes the EUV spectral index for radio-loud
quasars ($\alpha_{EUV}=-1.96$) and the larger assumes the radio-quiet
quasar index ($\alpha_{EUV}=-1.57$).  The Elvis et al.\ (1994)
template, which was used by HC, gives the substantially smaller
value ${\dot N}_{ph} = 0.2\times10^{58}$.  We think the
Telfer et al.\ spectrum is preferable but report results using
both spectra below.  The corresponding range
of values for $t_Q$ is 5--$7\times10^5$~yrs or $3\times10^6$~yrs (for
the Telfer and Elvis templates, respectively.)

Such a short lifetime for \qb\ would be surprising, since its high
luminosity implies a large black-hole mass which should take a
considerable length of time to assemble.  The $e$-folding timescale for
an accreting  black hole is
\begin{equation}
t_{acc} = 4\times10^7\,\hbox{yr}
\left(\epsilon \over 0.1\right)
\eta^{-1} \quad ,
\label{eqn-tacc}
\end{equation}
where $\epsilon$ is the radiative efficiency and $\eta$ is the ratio of the
quasar's luminosity to the Eddington luminosity (Haiman \& Loeb 2001).
This long natural timescale makes it unlikely that we will observe
quasars with ages as short as a few hundred thousand years unless the
radiative efficiency $\epsilon$ is very low, allowing the quasar to accrete
quietly and rapidly.

On the other hand, following HC's argument, if the quasar is lensed
then its true luminosity could be substantially smaller.  With a
plausible age of $t_Q\sim10^7$~yrs, \qb\ could be easily lensed by a
factor of 3 or more and still be able to ionize the IGM out to the
observed distance of 2.8~Mpc.

We note that for the sizes and ages derived for this object, the
Str\"omgren sphere is still in its early phase of rapid relativistic
expansion.  Even in that stage of evolution, however, Eqn.~(\ref{eqn-rs1})
accurately describes the observed size of the \ion{H}{2} region as
measured by absorption along the line-of-sight.  See the Appendix
for further details.

One more point that is worth making is that these discrepancies
become even greater if the IGM is not completely neutral at $z\sim6$.
The expected value of ${\dot N}_{ph}t_Q$ decreases in direct
proportion to the neutral fraction of the gas, so if the IGM at
$z=6$ has $\nhrat = 0.15$ as suggested by Cen (2003), the quasar
would have to be very highly magnified through lensing, in an
extremely overdense region of the IGM, or extraordinarily young in
order to produce an \ion{H}{2} region as small as that we observe.

What other explanations could there be for the small \ion{H}{2}
region observed in front of \qb?  One possibility is that the
redshift of \qb\ is underestimated, which would increase the
observed size of the ionized region.  If the quasar redshift is
6.41 instead of 6.37 (Willott, McLure \& Jarvis 2003),
the Str\"omgren radius increases to 4.7~Mpc.
That would increase the ${\dot N}_{ph}t_Q$ product in
Eqn.~(\ref{eqn-ndott}) by a factor of 4.8 to 0.31, which reduces
the discrepancy with the observed luminosity and expected lifetime.
However,
a redshift as high as 6.41 produces a very poor match between
the LBQS template and the observed spectrum (Fig.~\ref{fig-btemplate}),
so we consider a redshift that large unlikely.  Matching the
LBQS composite spectrum to the \ion{Mg}{2} emission line detected at
$2\mu$m by Willott et al.\ (2003) reveals that $z=6.37$ matches the
data about as well as $z=6.41$; in fact, from the LBQS spectrum,
which does include the offsets typically seen between the
redshifts of different emission lines, a redshift of 6.41
would appear to be an upper limit for \qb.  It will probably
require higher SNR infrared spectra of this object to resolve
the question of its true redshift.

Another possibility is that clumpy structure in the IGM conspires to
limit the size of the Str\"omgren sphere along the line of sight.
Certainly there is substantial IGM structure near the quasar, as it can
be seen in the narrow absorption lines on the blue edge of the
\lya\ emission line for wavelengths approaching the GP trough.  A
cloudy IGM can also modify the radiative transfer for ionizing
radiation so that the Str\"omgren sphere is not fully ionized (Cen \&
Haiman 2000; Fan et al.\ 2002), so that gas within the ionized region
contributes substantially to the observed \lya\ optical depth.

However, these effects ought to be relatively small for \ion{H}{2}
regions that are still in their very early phase of relativistic
expansion (see Appendix~A).  In that phase there is a surfeit of
ionizing photons, with far more photons present than are required to
ionize the neutral hydrogen.  The expansion rate of the ionization
front is limited by the light-travel time; recombinations are
completely negligible, and even dense gas clumps are fully ionized by
the copious radiation.

Stopping the ionization front with a dense cloud along the line
of sight is difficult for the same reason.  For example,
suppose there is a dense cloud in the IGM
at $z=6.32$.  If the cloud is optically thick to Lyman continuum
radiation, it will stop the Str\"omgren sphere expansion,
thus producing an anomalously small
ionized region.  A cloud dense enough to stop the ionization
front must have a density greater than
\begin{equation}
n_H = \left(
  {\dot N}_{ph} \over 4 \pi R_S^2 \alpha_B \ell
\right)^{1/2}
= 0.12\,{\rm cm}^{-3}\,
 \left(
   {{\dot N}_{ph} \over 10^{58}\,{\rm s}^{-1}}
   {{\rm kpc} \over \ell}
 \right)^{1/2}
 \left(R_S \over 2.8\,{\rm Mpc} \right)^{-1}
\quad ,
\label{eqn-clouddensity}
\end{equation}
where $\alpha_B$ is the case B recombination coefficient and
$\ell$ is the line-of-sight distance through the cloud.
If we assume the cloud is spherical, its mass is
\begin{equation}
M = 2\times10^6\,{\rm M}_\sun \,
\left( \ell \over {\rm kpc} \right)^{5/2}
\left( {\dot N}_{ph} \over 10^{58}\,{\rm s} \right)^{1/2}
\left( R_S \over 2.8\,{\rm Mpc} \right)^{-1} \quad .
\label{eqn-cloudmass}
\end{equation}
The cloud would have to be either dense or quite massive
to stop the quasar's ionization front.

Another possible explanation for the small size of the ionized region
is that the edge of the observed emission at $z=6.32$ does not indicate
the edge of the ionized gas.  Only a trace of neutral hydrogen is
needed to generate a large optical depth to \lya\ 
(eq.~\ref{eqn-taugp}).  Consequently, if the ionization fraction
in the \ion{H}{2} region drops below $\sim10^{-4}$, the IGM becomes
opaque.  This can occur if radiation transfer in the clumpy IGM reduces
the ionizing flux seen by the gas.  This also appears unlikely in the
Str\"omgren sphere's relativistic expansion phase, but more detailed
calculations are required to confirm that $z=6.32$ really marks the
position of the quasar's ionization front.

\subsection{Could \qa\ also be lensed?}

One might wonder whether the arguments presented in favor of a
foreground object lensing \qb\ could also be applied to \qa.  \qa\ does
show some absorption lines from intervening metal line systems
(Figs.~\ref{fig-spectra} and~\ref{fig-atemplate}), but none are as deep
as the strongest \ion{C}{4} absorption line system seen in \qb, nor do
we see any evidence for associated \lya\ emission.  The absorption
spectrum of \qa\ will be discussed further by Madau \& Bolte (in
preparation.)

There is no sign that continuum emission from an intervening galaxy
contaminates the \lya\ or \lyb\ GP troughs in \qa.  Indeed, those
troughs are very black, with residual levels even below those detected
in \qb\ (see Table~2).  From the flux limit in the \lya\ trough, we place
a limit of $m_{\rm AB} > 27.4$ on any intervening galaxy.
We can rule
out the presence of a lensing galaxy at $z<3.5$ with an absolute
AB magnitude brighter than $-20$ (at a rest wavelength of 2000~\AA).

Finally, Haiman \& Cen (2002) discussed the limits that can be placed
on any lensing of \qa\ from the size of its Str\"omgren sphere.  The
quasar has an observed $R_S=4.5$~Mpc at $z=6.28$, so we find $({\dot
N}_{ph}/10^{58}) (t_Q/10^7) = 0.25$.  The ionizing photon luminosity is
$0.13\times10^{58}\,\hbox{s}^{-1}$ or
0.61--$0.87\times10^{58}\,\hbox{s}^{-1}$ (from the Elvis and Telfer
templates), leading to age estimates of $2\times10^7$~yrs or
3--$4\times10^6$~yrs.  These age estimates appear plausible,
although the higher ionizing luminosities using the Telfer et
al.\ template do leave more room for lensing than was found by HC.

\subsection{Summary}

Our proposal gives a consistent picture for \qb: the GP troughs of both
\lya\ and \lyb\ are likely to be contaminated with light from an
intervening galaxy at $z\sim5$.  The existence of this galaxy is
supported by the detection of \lya\ emission, \ion{C}{4} absorption,
and continuum emission.  There could be several such intervening
systems, since both the emission and absorption are complex with
multiple components spread over $\sim1000$~km/s.  The intervening
system amplifies the quasar's light via gravitational lensing,
thereby enhancing the likelihood of discovery for the quasar.  The
possibility of strong lensing is supported by the relatively small \ion{H}{2}
region created by the quasar's ionizing radiation, which indicates that
the quasar is probably considerably less luminous than its apparent
brightness would indicate.

There may be some way to reconcile the observations of \qb\ presented
here with a model based purely on IGM absorption.  No such explanation
is obvious to us, though there are so many excellent theoreticians
who are interested in this problem that there will doubtless be many
good ideas proposed shortly after the publication of these results!
From the observational side, clearly an HST image of \qb\ would be very
interesting.  It could reveal the presence of multiple images (or set
strong limits on their absence.) An HST image taken using a narrow-band
filter in the \lyb\ trough would directly determine whether the peaks
seen at 7205 and 7236~\AA\ are extended and offset (as expected if they are
\lya\ emission from an intervening galaxy) or are point-like and
coincident with the quasar (as expected if they are the result of
``leaks'' in the IGM absorption.)

If the intervening galaxy suggestion turns out to be correct, then our
observations of \qa\ remain the single best measurement of the
transmission of the IGM at $z>6$.  There is no light detected in the
\lya\ or \lyb\ troughs for this object at a very low level.  On the
other hand, if \qb\ is shown {\sl not} to have an intervening system
contaminating its GP troughs, then the IGM toward it is in fact quite
transparent, with both numerous high-ionization holes and a significant
transmission across the whole \lya\ and \lyb\ troughs.  The resolution
of these questions will have to wait for the discovery of additional
$z>6$ quasars.

\acknowledgments

Many thanks to Michael Bolte and Piero Madau for sharing their ESI
exposures of \qa\ with us.  We appreciate a helpful conversation
with James Rhoads and Sangeeta Malhotra.
RLW thanks Holland Ford and the Johns
Hopkins University Astronomy \& Physics Department for their
hospitality during a sabbatical, when much of this work was carried
out, and appreciates continuing support from the Space Telescope
Science Institute.  RHB acknowledges support from NSF grant
AST-00-98355 and the Institute of Geophysics and Planetary Physics
(operated under the auspices of the U.~S.\ Department of Energy by
Lawrence Livermore National Laboratory under contract No.
W-7405-Eng-48).  XF acknowledges support from the University of
Arizona and an Alfred P.~Sloan Research Fellowship.
MAS acknowledges support from NSF
grant AST-00-71091.

\appendix

\section{Appendix: Early Evolution of a Str\"omgren Sphere}
\label{section-appendix}

The early evolution of the ionized region around a quasar
is marked by a period of very rapid expansion, with the
ionizing front moving out at nearly the speed of light.
In that case the finite light travel time across the
Str\"omgren sphere cannot be ignored.  However, the
size of the sphere as inferred from observations of
absorption along the line-of-sight to the quasar turns
out to have exactly the
same evolution with time as one derives under the assumption
of an infinite speed of light.  This appendix derives the
evolution of the size both as seen in the frame of the
ionizing source and as seen through the line-of-sight
absorption.

We assume that the mean free path for ionizing photons in the
neutral gas is very short and that recombinations and the expansion
of the universe can be ignored.  These are good approximations:
the mean free path for 1 Rydberg photons is only 1~kpc at $z=6$
and declines as $(1+z)^3$.  Cosmological simulations indicate that
the gas clumping factor $C = \langle n_H^2\rangle/\langle n_H\rangle^2$
is likely to be well below the value $C\sim100$ required to make
recombinations a significant factor (Gnedin \& Ostriker 1997; Madau
\& Rees 2000; Cen \& Haiman 2000).

\begin{figure*}
\epsscale{0.45}
\plotone{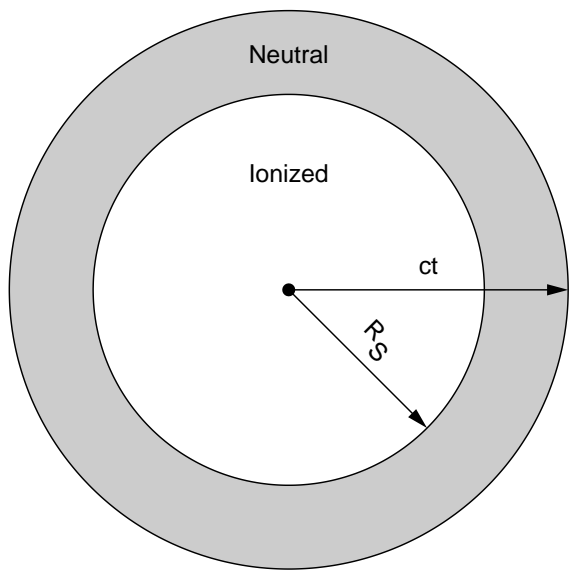}
\epsscale{1.0}
\caption{
Schematic diagram of early evolution of a Str\"omgren sphere.
The outer circle of radius $ct$ marks the extent (in the absence
of absorption) of the ionizing
radiation at a time $t$ after the source turns on.  The inner
circle of radius $R_S$ is the ionized Str\"omgren sphere.
The number of photons in the shaded volume is equal to the
number of hydrogen atoms in the ionized volume.
}
\label{fig-rsdiag}
\end{figure*}

With these assumptions, ionizing photons propagate freely to the
edge of the ionized region where they are immediately absorbed by
the neutral gas.  Figure~\ref{fig-rsdiag} shows the relevant geometry.
The photons that would have traveled beyond the Str\"omgren radius
$R_S$ (the shaded region)
have been absorbed by neutral hydrogen atoms within $R_S$.
The size of the \ion{H}{2} region is determined by a balance between
the number of photons in the outer shell and the number of atoms
within the ionized sphere:
\begin{equation}
{\dot N}_{ph} \left(t - {R_S \over c} \right) =
{4 \pi R_S^3 n_H \over 3}  \quad ,
\label{eqn-balance}
\end{equation}
where ${\dot N}_{ph}$ is the ionizing photon luminosity
and $t$ is the age of the source.  This cubic equation
can be analytically solved to give
\begin{equation}
R_S(t) = { 3y \over y^2+y+1 } ct \quad ,
\label{eqn-rs}
\end{equation}
where
\begin{equation}
y = \left({3^{3/2} \over 2} {t \over t_c} +
\sqrt{{27 \over 4} \left(t \over t_c\right)^2+1} \right)^{2/3}
\label{eqn-y}
\end{equation}
and the time $t_c$, which roughly marks the end of the relativistic
expansion period, is
\begin{equation}
t_c = \left( 3 {\dot N}_{ph} \over 4 \pi n_Hc^3 \right)^{1/2} 
= 1.2\times10^7 \, \hbox{yr} \,
\left({\dot N}_{ph} \over 10^{57}\,\hbox{s}^{-1}\right)^{1/2} \,
\left(1+z_Q \over 7\right)^{-3/2} \,
\left({\nhratover} {\Omega_b h^2 \over 0.0196}\right)^{-1/2}
\quad ,
\label{eqn-tc}
\end{equation}
where $z_Q$ is the quasar redshift.
Here we have explicitly included the dependence on the
neutral fraction $\nhrat$ and the cosmological parameters.

\begin{figure*}
\epsscale{0.6}
\plotone{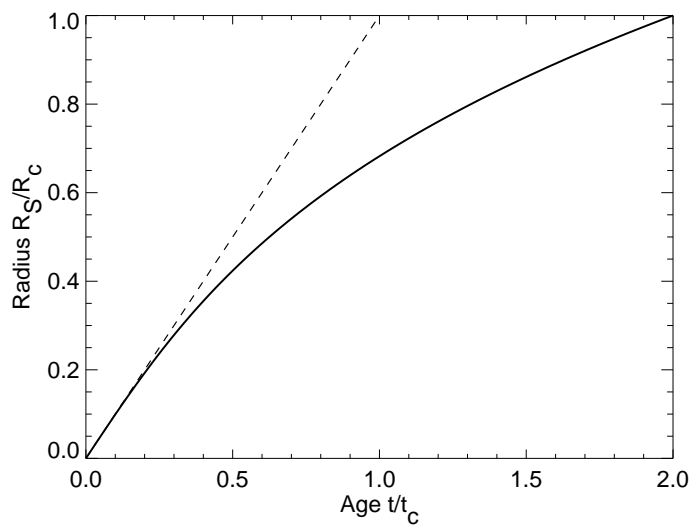}
\epsscale{1.0}
\caption{
Size versus time for early Str\"omgren sphere expansion.
The time is scaled by $t_c$, the duration of relativistic
expansion (see text for details) and the radius by $R_c = ct_c$.
The dashed line is $R=ct$.  At late times the expansion is
$R \sim t^{1/3}$.
}
\label{fig-rsevol1}
\end{figure*}

The expansion law in Eqn.~(\ref{eqn-rs}) is shown in
Figure~\ref{fig-rsevol1}.  After an initial period where the
ionization front moves out at nearly the speed of light, the
expansion slows and approaches the standard $t^{1/3}$ law (Madau
\& Rees 2000, Cen \& Haiman 2000).

This is not, however, the expansion law that is observed via
absorption along the line-of-sight.  When we observe \lya\ absorption
from the neutral gas just beyond the edge of the \ion{H}{2} region,
the \lya\ photons being detected were emitted only a short time
after the first ionizing photons.  Specifically, if the \lya\
photons cross the Str\"omgren boundary at a radius $R_S$, they were
emitted when the age of the quasar was only $t' = t-R_S/c$ (which is the
time for light to cross the outer, shaded shell in Fig.~\ref{fig-rsdiag}).
The observed Str\"omgren evolution, as inferred from absorption,
is not $R_S(t)$ versus $t$ but rather $R_S(t)$ versus $t'$.
With some algebraic manipulation of the analytical expressions
for $R_S$ and $t'$, we can express $R_S$ directly in terms of
the observed (apparent) time $t'$ as
\begin{equation}
R_S = R_c \left( t' \over t_c \right)^{1/3} \quad ,
\label{eqn-rsobs}
\end{equation}
where $R_c = ct_c$.

This is a surprising result (at least, it surprised us!)  The
expansion law for the observed radius is exactly the same as
the expansion law derived if one completely ignores light-travel time
effects (eq.~\ref{eqn-rs1}).
In the frame of the quasar the ionization front's expansion
velocity is limited by the speed of light, but for the observer
there is an initial period of superluminal expansion.
The speed of light drops out of the solution along
the line-of-sight because the delay required to allow light 
to travel from the source to the edge of the \ion{H}{2} region
is exactly compensated by the speedup that results from that
edge being closer to the observer, allowing light originating
there to reach the observer sooner than light from the quasar core.

Note that the size of the observed \ion{H}{2} region in
Eqn.~(\ref{eqn-rsobs}) may be larger than the ionized region that exists
at the end of the quasar's lifetime.  For example, suppose a quasar
shines for a period $t_c$ and then shuts down.  According to
Eqn.~(\ref{eqn-rs}), at $t_c$ the radius of the \ion{H}{2} region is
$0.7\,R_c$.  But the observer measures a radius $R_c$
(eq.~\ref{eqn-rsobs}).  What is happening in this case is that the
\ion{H}{2} region continues to expand after the quasar shuts off
until it reaches a maximum size $R_c$ at $t=2t_c$.  At that point
the inner sphere in Figure~\ref{fig-rsdiag} is completely empty of
ionizing photons and so the expansion stops.  The observer is
detecting \lya\ photons that were emitted just at the end of the
quasar's lifetime, but by the time those photons cross the ionization
front it has expanded to its maximum size.

Our conclusion is that when all the light travel time effects are
taken into account, Eqn.~(\ref{eqn-rsobs}) is the correct formula
for computing the age of the quasar given the luminosity.  Ages
determined using that formula are, in fact, the age of the quasar
at the time it emitted the photons being detected, and so such ages
are direct measurements of the minimum quasar lifetime even
during the superluminal expansion phase.

\end{document}